\documentclass[a4paper,11pt]{article}
\usepackage{pos, tikz-feynman}

\renewcommand{\logo}{\relax}

\title{QED corrections to meson masses}

\author*{Joshua Swaim}

\affiliation{University of Connecticut,\\
  Storrs, Connecticut, United States}

\emailAdd{joshua.swaim@uconn.edu}

\abstract{We present our progress on calculating leading-order QED corrections to meson masses and bare quark masses. As lattice QCD calculations become more precise, these QED corrections are becoming more important. However, one of the challenges in adding QED effects to QCD calculations is avoiding power-law suppressed finite-volume effects. By using the recently introduced infinite-volume reconstruction method for QED, we are able to avoid this problem and perform calculations with exponentially-suppressed finite-volume effects.}

\FullConference{The 40th International Symposium on Lattice Field Theory (Lattice 2023)\\
July 31st - August 4th, 2023\\
Fermi National Accelerator Laboratory\\}

\begin{document}
\maketitle

\section{Introduction}
As lattice QCD continues to progress, precision calculations increasingly need to account for QED and isospin breaking corrections \cite{FLAG}. For example, QED corrections are needed for precise calculations of the muon anomalous magnetic moment \cite{Giusti_HVP_2019, MILC_HVP_2019, Biloshytskyi_HVP_2022}. These corrections are important not only because they directly affect observables, but also because they change the definition of the physical point \cite{DiCarlo_matching_2019, Boyle_matching_2022}. Since meson masses are frequently used when defining the physical point, it is important to understand QED corrections to meson masses. Several groups have recent work on this topic \cite{Davoudi_2014, Horsley_2015, Horsley2_2015, ETM_2017, Boyle_2017, MILC_2018, Risch_QEDL_Open_Bound_2018, Hatton_heavy_mesons_2020, RCstar_2022, RM123_2022}.

Introducing QED to lattice calculations is challenging. One problem is that QED contains unconfined massless degrees of freedom: the photons.  Photon propagators, unlike propagators for massive particles, do not decay exponentially at large distances. This means that introducing QED naively leads to large finite-volume errors. Furthermore, combining QED with periodic boundary conditions leads to technical complications. There are several formulations which address these difficulties in various ways, such as QED${}_\text{TL}$ \cite{Duncan_QED_TL}, QED$_\text{L}$ \cite{Hayakawa_QED_L}, QED with massive photons \cite{Endres_QED_M}, and QED with C${}^*$ boundary-conditions \cite{Lucini_QED_star}.

In 2018, a new method was introduced that avoids the challenges posed both by volume and periodic boundary conditions. This method is called the infinite-volume reconstruction method \cite{Feng_power-law_2018}. In this method, QED corrections are calculated semi-analytically in infinite volume. Contributions from outside the lattice volume are reconstructed with exponentially-suppressed systematic errors. In this paper, we present our progress on using this method to calculate QED corrections to meson and quark masses.

In section \ref{QCD+QED}, we discuss how QED corrections are calculated based on QCD correlation functions. In section \ref{inf_reconstruction}, we explain the infinite-volume reconstruction method. In section \ref{meson_masses}, we demonstrate that this method works for calculating QED corrections to meson masses. Finally, in section \ref{quark_masses}, we briefly discuss our work so far in extracting the QED corrections to the quark mass renormalization constants.

\subsection{Adding QED Corrections to QCD} \label{QCD+QED} 
\begin{figure}
	\centering
	\begin{tikzpicture}
		\begin{feynman}
			\vertex (x) at (0,0.5);
			\vertex (y) at (6,0.5);
			\vertex (a) at (2,0.5);
			\vertex (b) at (4,0.5);
			\diagram* {
				(x) -- (a),
				(a) -- (b),
				(a) -- [quarter left, photon] (b),
				(y) -- (b)
			};
		\end{feynman}
	\end{tikzpicture}
	\caption{A diagramatic representation of the leading-order QED correction to a hadronic propagator. The straight lines represent hadronic propagators computed non-perturbatively using only QCD, and the other line represents a free photon propagator. The points where the lines connect represent current insertions.}
	\label{feyn_photon}
\end{figure}
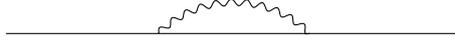

In the infinite-volume reconstruction method \cite{Feng_power-law_2018}, QED is introduced perturbatively. By expanding the path integral in the electric charge $e$, we get
\begin{align} \label{QED_pert}
	\nonumber \langle \mathcal{O}(T) \mathcal{O}(-T)&\rangle_{\text{QCD}+\text{QED}} = \langle\mathcal{O}(T) \mathcal{O}(-T)\rangle_\text{QCD} \\
	&+\frac{e^2}{2}\int d^4xd^4y \langle \mathcal{O}(T) J_\mu(x) J_\nu(y)\mathcal{O}(-T)\rangle_\text{QCD} S_{\mu\nu}(x-y)+\mathcal{O}(e^4),
\end{align}
where $S_{\mu\nu}(x-y)$ is the photon propagator. $\langle\rangle_\text{QCD+QED}$ represents the vacuum expectation value of operators in the full theory of QCD+QED, while $\langle\rangle_\text{QCD}$ represents the vacuum expectation value of operators computed using only QCD. If $\mathcal{O}$ is an operator that creates a hadron, the order $e^2$ correction can be represented diagrammatically, as shown in Figure \ref{feyn_photon}. Based on equation \ref{QED_pert}, the leading-order QED correction to the mass of a hadron is given by 
\begin{equation} \label{mass_correction}
	\Delta m = \frac{e^2}{2}\int d^4x \mathcal{H}_{\mu\nu}(x)S_{\mu\nu}(x),
\end{equation}
where, on the lattice, 
\begin{equation}
	\mathcal{H}_{\mu\nu}(x)=L^3\frac{\langle \mathcal{O}(t+T)J_\mu(x)J_\nu(0)\mathcal{O}(-T)\rangle_\text{QCD}}{\langle\mathcal{O}(t+T)\mathcal{O}(-T)\rangle_\text{QCD}},
\end{equation}
and $\mathcal{O}$ is an operator that creates the desired hadronic state. In infinite volume, assuming for example that $\mathcal{O}$ creates a pion, this definition would correspond to 
\begin{equation}
	\mathcal{H}_{\mu\nu}(x)=\frac{1}{2m}\langle\pi|J_\mu(x)J_\nu(0)|\pi\rangle_\text{QCD}.
\end{equation}

\subsection{The Infinite-Volume Reconstruction Method} \label{inf_reconstruction}
We could use equation \ref{mass_correction} to estimate $\Delta m$ by evaluating $\mathcal{H}_{\mu\nu}(x)$ using lattice QCD and replacing the integral over space with a sum over the lattice volume. However, this would result in large finite-volume errors. To see this, note that when $t>>|\vec{x}|$, $\mathcal{H}_{\mu\nu}(x)$ is order $1$, even at large distances. Similarly, the photon propagator $S_{\mu\nu}(t,\vec{x})$ is only power-law (not exponentially) suppressed at large $t$ because the photon is massless. Therefore, the finite-volume errors resulting from evaluating equation \ref{mass_correction} on the lattice will only be power-law suppressed.

To get exponentially-suppressed finite-volume effects, we can reconstruct the large-distance contributions to the integral using the infinite-volume reconstruction method \cite{Feng_power-law_2018}. At large $|x|$, $\mathcal{H}_{\mu\nu}(x)$ is dominated by contributions from the lowest energy states.
We choose some cutoff time $t_s$ that is large enough for $\mathcal{H}_{\mu\nu}(t_s,\vec{x})$ to be dominated by the single-meson intermediate states, but small enough that we don't need to worry about around-the-world effects.
Then we can reconstruct $\mathcal{H}_{\mu\nu}(t,\vec{x})$ for $t>t_s$ using
\begin{equation}
	\mathcal{H}_{\mu\nu}(t,\vec{x}')\approx\int d^3\vec{x}\mathcal{H}_{\mu\nu}(t_s,\vec{x})\int\frac{d^3\vec{p}}{(2\pi)^3}e^{-i\vec{p}\cdot(\vec{x}'-\vec{x})}e^{-(E_{n,\vec{p}}-m_\pi)(t-t_s)},
\end{equation}
with corrections to this formula exponentially suppressed (see \cite{Feng_power-law_2018} for more details).

\section{Results}
\subsection{Calculating Meson Masses} \label{meson_masses}
\begin{figure}
	\centering
	\includegraphics[width=350pt]{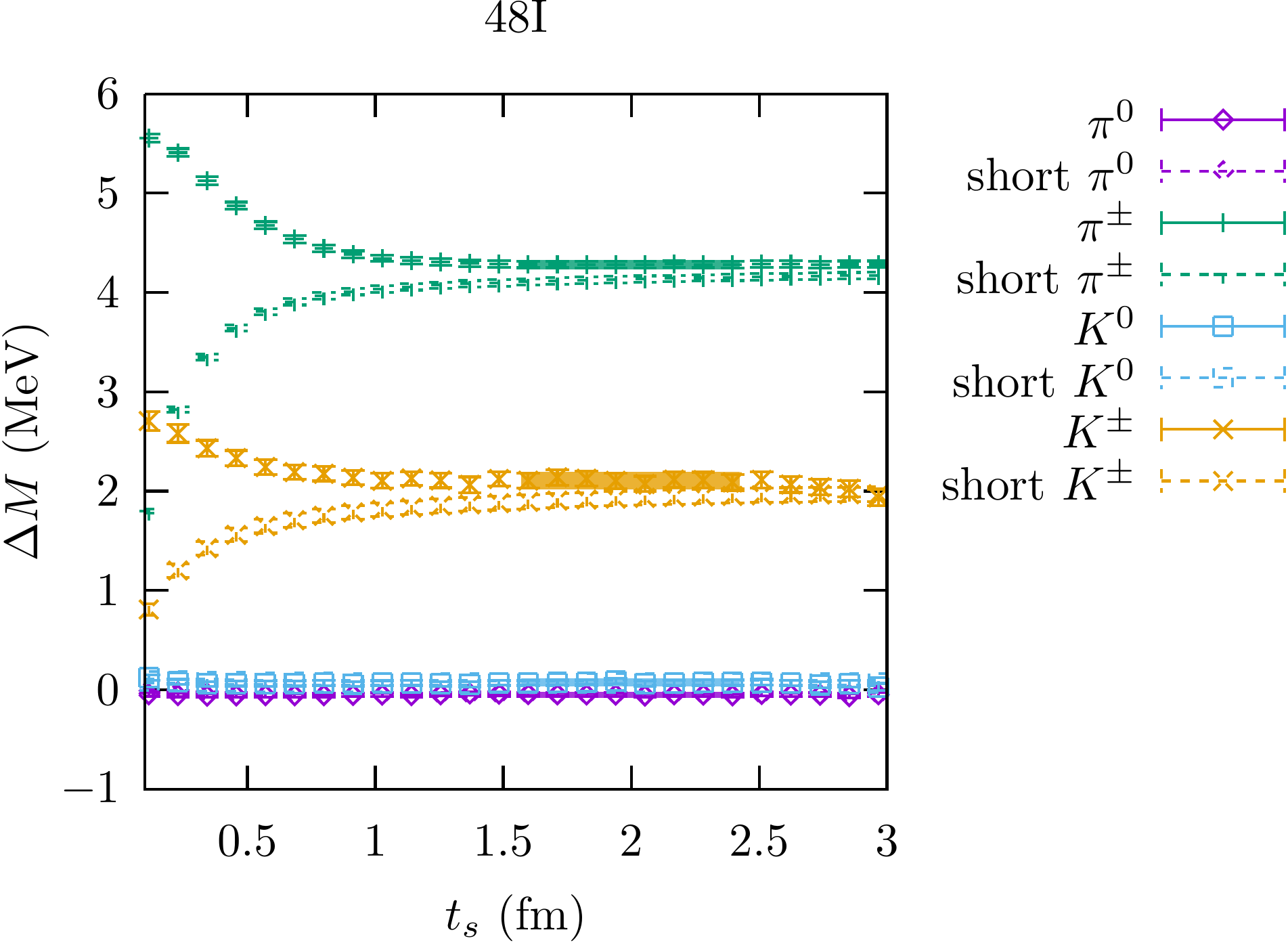}
	\includegraphics[width=350pt]{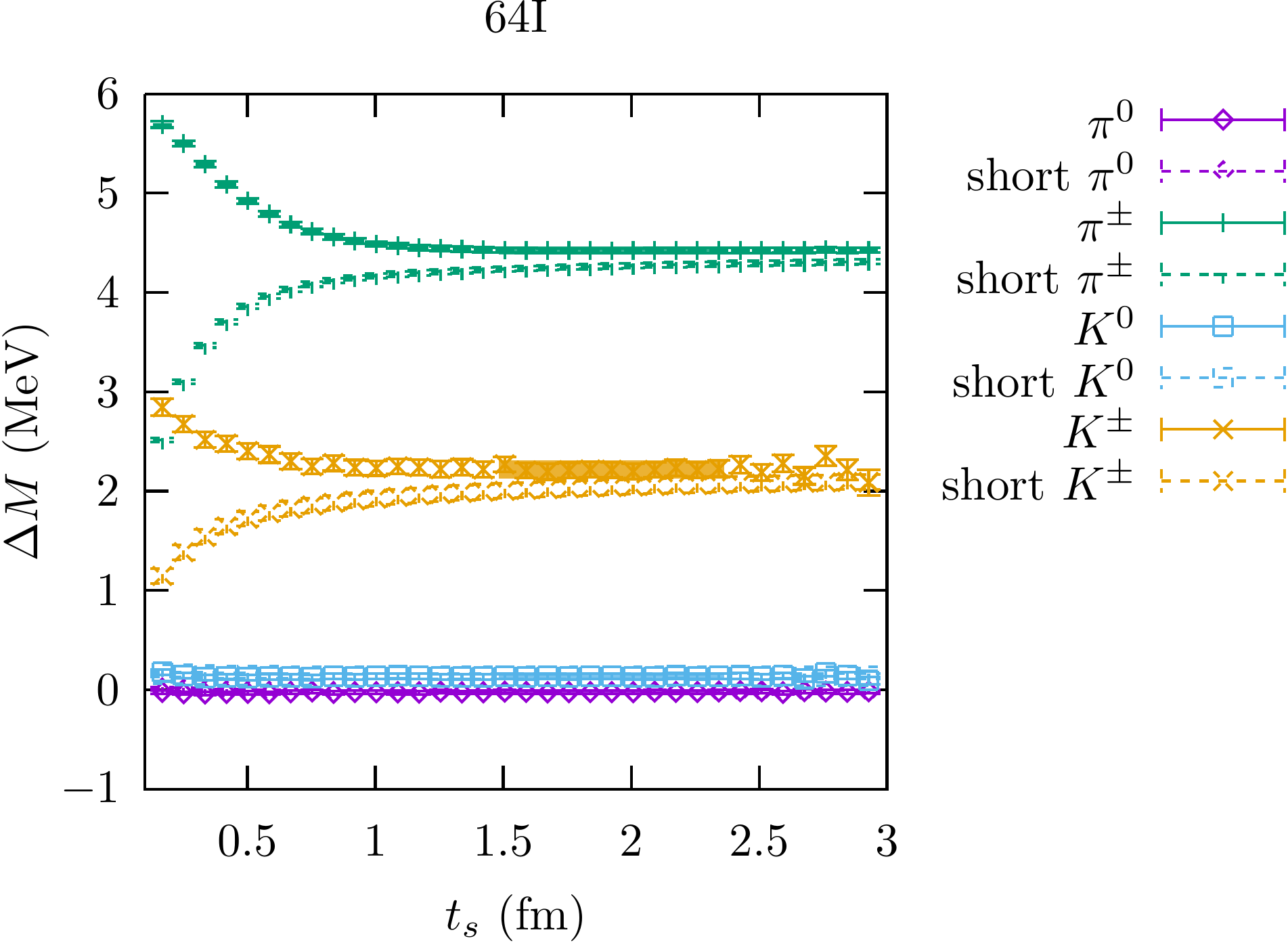}
	\caption{$\Delta M$ versus $t_s$ on a $48^3\times 96$ lattice (top) and $64^3\times 128$ lattice (bottom) on ensembles from RBC/UKQCD \cite{RBC_Ensembles_2016} using Iwasaki gauge action and domain-wall fermions. "Short" means including only $|t|<t_s$ contributions.}
	\label{delta_m_plot}
\end{figure}

In figure \ref{delta_m_plot}, we show our calculated $\Delta m$ based on equation \ref{mass_correction} as a function of the cutoff time $t_s$. We show both the results with the infinite-volume reconstruction and the "short" results where the integral is simply cutoff at time $t_s$ and no infinite-volume reconstruction is performed. There is a plateau region (highlighted in the plots) where the calculated mass correction does not depend strongly on $t_s$. This indicates that we can indeed choose $t_s$ sufficiently large that the infinite-volume reconstruction works well.

\subsection{Quark Masses and Renormalization} \label{quark_masses}
In QCD, the quark masses renormalize by a multiplicative constant
\begin{equation}
	m_f^{\overline{\text{MS}},\text{QCD}} = Z_{m}m_f,
\end{equation}
where $m_f$ is the bare quark mass of flavor $f$ and $m_f^{\overline{\text{MS}},\text{QCD}}$ is the renormalized quark mass in QCD (without QED corrections). Adding QED to the theory introduces additional divergences. Therefore, the renormalization constant is modified. We define $Z_\text{QED}$ by
\begin{equation}
	m_f^{\overline{\text{MS}}} = Z_m(1+e_f^2Z_\text{QED})m_f,
\end{equation}
where $m_f^{\overline{\text{MS}}}$ is the renormalized quark mass taking both QCD and QED into account.

To get $Z_\text{QED}$, we note that hadron masses are renormalization-invariant by definition. To see how we can use this fact, suppose that we could calculate the shift $\Delta m_H$ in the mass of a hadron $H$ caused by making a small change $\Delta m_f^{\overline{\text{MS}}}$ in an $\overline{\text{MS}}$ quark mass. We could then figure out the shift $\Delta m_f$ in the bare lattice quark mass that would be required to produce the same $\Delta m_H$. By comparing the shift in $\overline{\text{MS}}$ quark mass to the equivalent shift in bare quark mass, we could determine the renormalization constant.

To the leading order, the change in hadron mass, $m_H$, due to a change in the quark mass $m_f$ and introducing an electric charge $e$ is
$$
\Delta m_{H}=\frac{e^2}{2}\int d^4 x \mathcal{H}_{\mu\nu}(x)S_{\mu\nu}(x)+\Delta m_{f}\mathcal{H}^{\text{3pt}}_{f},
$$
where $\mathcal{H}$ is the four-point function, and (in the lattice normalization) $$\mathcal{H}_f^\text{3pt}=L^3\frac{\langle{\mathcal{O}_H(T)\bar\psi_f(0)\psi_f(0)\mathcal{O}_H(-T)}\rangle}{\langle{\mathcal{O}_H(T) \mathcal{O}_H(-T)}\rangle}.$$ In $\overline{\text{MS}}$, we can calculate the divergent part of the integral using the operator product expansion \cite{Wilson_OPE_1972}. We can compare this with the small-distance (high-momentum) contribution to this integral from the lattice.

\section{Conclusion}
We demonstrated that the infinite-volume reconstruction method can be used to get QED corrections to meson masses with exponentially-suppressed finite-volume effects. To get final results, we still need to perform a continuum extrapolation and choose a scheme to match our simulation parameters to the physical world. To get quark mass corrections, we need to determine QED corrections to the renormalization constants.

\acknowledgments
\noindent J.S. acknowledges the support of DOE Office of Science Early Career Award DE-SC0021147 and DOE grant DE-SC0010339.

\bibliographystyle{JHEP}
\bibliography{bibliography}

\begin{thebibliography}{23}
\providecommand{\natexlab}[1]{#1}
\providecommand{\url}[1]{\texttt{#1}}
\expandafter\ifx\csname urlstyle\endcsname\relax
  \providecommand{\doi}[1]{doi: #1}\else
  \providecommand{\doi}{doi: \begingroup \urlstyle{rm}\Url}\fi

\bibitem[Aoki et~al.(2022)]{FLAG}
Y.~Aoki et~al.
\newblock {FLAG Review 2021}.
\newblock \emph{Eur. Phys. J. C}, 82\penalty0 (10):\penalty0 869, 2022.
\newblock \doi{10.1140/epjc/s10052-022-10536-1}.

\bibitem[Giusti et~al.(2019)Giusti, Lubicz, Martinelli, Sanfilippo, and
  Simula]{Giusti_HVP_2019}
D.~Giusti, V.~Lubicz, G.~Martinelli, F.~Sanfilippo, and S.~Simula.
\newblock {Electromagnetic and strong isospin-breaking corrections to the muon
  $g - 2$ from Lattice QCD+QED}.
\newblock \emph{Phys. Rev. D}, 99\penalty0 (11):\penalty0 114502, 2019.
\newblock \doi{10.1103/PhysRevD.99.114502}.

\bibitem[Davies et~al.(2020)]{MILC_HVP_2019}
C.~T.~H. Davies et~al.
\newblock {Hadronic-vacuum-polarization contribution to the
  muon\textquoteright{}s anomalous magnetic moment from four-flavor lattice
  QCD}.
\newblock \emph{Phys. Rev. D}, 101\penalty0 (3):\penalty0 034512, 2020.
\newblock \doi{10.1103/PhysRevD.101.034512}.

\bibitem[Biloshytskyi et~al.(2023)Biloshytskyi, Chao, G\'erardin, Green,
  Hagelstein, Meyer, Parrino, and Pascalutsa]{Biloshytskyi_HVP_2022}
Volodymyr Biloshytskyi, En-Hung Chao, Antoine G\'erardin, Jeremy~R. Green,
  Franziska Hagelstein, Harvey~B. Meyer, Julian Parrino, and Vladimir
  Pascalutsa.
\newblock {Forward light-by-light scattering and electromagnetic correction to
  hadronic vacuum polarization}.
\newblock \emph{JHEP}, 03:\penalty0 194, 2023.
\newblock \doi{10.1007/JHEP03(2023)194}.

\bibitem[Di~Carlo et~al.(2019)Di~Carlo, Giusti, Lubicz, Martinelli, Sachrajda,
  Sanfilippo, Simula, and Tantalo]{DiCarlo_matching_2019}
M.~Di~Carlo, D.~Giusti, V.~Lubicz, G.~Martinelli, C.~T. Sachrajda,
  F.~Sanfilippo, S.~Simula, and N.~Tantalo.
\newblock {Light-meson leptonic decay rates in lattice QCD+QED}.
\newblock \emph{Phys. Rev. D}, 100\penalty0 (3):\penalty0 034514, 2019.
\newblock \doi{10.1103/PhysRevD.100.034514}.

\bibitem[Boyle et~al.(2023)]{Boyle_matching_2022}
Peter Boyle et~al.
\newblock {Isospin-breaking corrections to light-meson leptonic decays from
  lattice simulations at physical quark masses}.
\newblock \emph{JHEP}, 02:\penalty0 242, 2023.
\newblock \doi{10.1007/JHEP02(2023)242}.

\bibitem[Davoudi and Savage(2014)]{Davoudi_2014}
Zohreh Davoudi and Martin~J. Savage.
\newblock {Finite-Volume Electromagnetic Corrections to the Masses of Mesons,
  Baryons and Nuclei}.
\newblock \emph{Phys. Rev. D}, 90\penalty0 (5):\penalty0 054503, 2014.
\newblock \doi{10.1103/PhysRevD.90.054503}.

\bibitem[Horsley et~al.(2016{\natexlab{a}})]{Horsley_2015}
R.~Horsley et~al.
\newblock {QED effects in the pseudoscalar meson sector}.
\newblock \emph{JHEP}, 04:\penalty0 093, 2016{\natexlab{a}}.
\newblock \doi{10.1007/JHEP04(2016)093}.

\bibitem[Horsley et~al.(2016{\natexlab{b}})]{Horsley2_2015}
R.~Horsley et~al.
\newblock {Isospin splittings of meson and baryon masses from three-flavor
  lattice QCD + QED}.
\newblock \emph{J. Phys. G}, 43\penalty0 (10):\penalty0 10LT02,
  2016{\natexlab{b}}.
\newblock \doi{10.1088/0954-3899/43/10/10LT02}.

\bibitem[Giusti et~al.(2017)Giusti, Lubicz, Tarantino, Martinelli, Sanfilippo,
  Simula, and Tantalo]{ETM_2017}
D.~Giusti, V.~Lubicz, C.~Tarantino, G.~Martinelli, F.~Sanfilippo, S.~Simula,
  and N.~Tantalo.
\newblock {Leading isospin-breaking corrections to pion, kaon and charmed-meson
  masses with Twisted-Mass fermions}.
\newblock \emph{Phys. Rev. D}, 95\penalty0 (11):\penalty0 114504, 2017.
\newblock \doi{10.1103/PhysRevD.95.114504}.

\bibitem[Boyle et~al.(2017)Boyle, G\"ulpers, Harrison, J\"uttner, Lehner,
  Portelli, and Sachrajda]{Boyle_2017}
P.~Boyle, V.~G\"ulpers, J.~Harrison, A.~J\"uttner, C.~Lehner, A.~Portelli, and
  C.~T. Sachrajda.
\newblock {Isospin breaking corrections to meson masses and the hadronic vacuum
  polarization: a comparative study}.
\newblock \emph{JHEP}, 09:\penalty0 153, 2017.
\newblock \doi{10.1007/JHEP09(2017)153}.

\bibitem[Basak et~al.(2019)]{MILC_2018}
S.~Basak et~al.
\newblock {Lattice computation of the electromagnetic contributions to kaon and
  pion masses}.
\newblock \emph{Phys. Rev. D}, 99\penalty0 (3):\penalty0 034503, 2019.
\newblock \doi{10.1103/PhysRevD.99.034503}.

\bibitem[Risch and Wittig(2018)]{Risch_QEDL_Open_Bound_2018}
Andreas Risch and Hartmut Wittig.
\newblock {Towards leading isospin breaking effects in mesonic masses with open
  boundaries}.
\newblock \emph{PoS}, LATTICE2018:\penalty0 059, 2018.
\newblock \doi{10.22323/1.334.0059}.

\bibitem[Hatton et~al.(2020)Hatton, Davies, and
  Lepage]{Hatton_heavy_mesons_2020}
D.~Hatton, C.~T.~H. Davies, and G.~P. Lepage.
\newblock {QED interaction effects on heavy meson masses from lattice QCD+QED}.
\newblock \emph{Phys. Rev. D}, 102\penalty0 (9):\penalty0 094514, 2020.
\newblock \doi{10.1103/PhysRevD.102.094514}.

\bibitem[Bushnaq et~al.(2023)Bushnaq, Campos, Catillo, Cotellucci, Dale,
  Fritzsch, L\"ucke, Krsti\'c~Marinkovi\'c, Patella, and Tantalo]{RCstar_2022}
Lucius Bushnaq, Isabel Campos, Marco Catillo, Alessandro Cotellucci, Madeleine
  Dale, Patrick Fritzsch, Jens L\"ucke, Marina Krsti\'c~Marinkovi\'c, Agostino
  Patella, and Nazario Tantalo.
\newblock {First results on QCD+QED with C* boundary conditions}.
\newblock \emph{JHEP}, 03:\penalty0 012, 2023.
\newblock \doi{10.1007/JHEP03(2023)012}.

\bibitem[Frezzotti et~al.(2022)Frezzotti, Gagliardi, Lubicz, Martinelli,
  Sanfilippo, and Simula]{RM123_2022}
R.~Frezzotti, G.~Gagliardi, V.~Lubicz, G.~Martinelli, F.~Sanfilippo, and
  S.~Simula.
\newblock {Lattice calculation of the pion mass difference
  M\ensuremath{\pi}+-M\ensuremath{\pi}0 at order O(\ensuremath{\alpha}em)}.
\newblock \emph{Phys. Rev. D}, 106\penalty0 (1):\penalty0 014502, 2022.
\newblock \doi{10.1103/PhysRevD.106.014502}.

\bibitem[Duncan et~al.(1996)Duncan, Eichten, and Thacker]{Duncan_QED_TL}
A.~Duncan, E.~Eichten, and H.~Thacker.
\newblock {Electromagnetic splittings and light quark masses in lattice QCD}.
\newblock \emph{Phys. Rev. Lett.}, 76:\penalty0 3894--3897, 1996.
\newblock \doi{10.1103/PhysRevLett.76.3894}.

\bibitem[Hayakawa and Uno(2008)]{Hayakawa_QED_L}
Masashi Hayakawa and Shunpei Uno.
\newblock {QED in finite volume and finite size scaling effect on
  electromagnetic properties of hadrons}.
\newblock \emph{Prog. Theor. Phys.}, 120:\penalty0 413--441, 2008.
\newblock \doi{10.1143/PTP.120.413}.

\bibitem[Endres et~al.(2016)Endres, Shindler, Tiburzi, and
  Walker-Loud]{Endres_QED_M}
Michael~G. Endres, Andrea Shindler, Brian~C. Tiburzi, and Andre Walker-Loud.
\newblock {Massive photons: an infrared regularization scheme for lattice
  QCD+QED}.
\newblock \emph{Phys. Rev. Lett.}, 117\penalty0 (7):\penalty0 072002, 2016.
\newblock \doi{10.1103/PhysRevLett.117.072002}.

\bibitem[Lucini et~al.(2016)Lucini, Patella, Ramos, and
  Tantalo]{Lucini_QED_star}
Biagio Lucini, Agostino Patella, Alberto Ramos, and Nazario Tantalo.
\newblock {Charged hadrons in local finite-volume QED+QCD with C*
  boundary conditions}.
\newblock \emph{JHEP}, 02:\penalty0 076, 2016.
\newblock \doi{10.1007/JHEP02(2016)076}.

\bibitem[Feng and Jin(2019)]{Feng_power-law_2018}
Xu~Feng and Luchang Jin.
\newblock {QED self energies from lattice QCD without power-law finite-volume
  errors}.
\newblock \emph{Phys. Rev. D}, 100\penalty0 (9):\penalty0 094509, 2019.
\newblock \doi{10.1103/PhysRevD.100.094509}.

\bibitem[Blum et~al.(2016)]{RBC_Ensembles_2016}
T.~Blum et~al.
\newblock {Domain wall QCD with physical quark masses}.
\newblock \emph{Phys. Rev. D}, 93\penalty0 (7):\penalty0 074505, 2016.
\newblock \doi{10.1103/PhysRevD.93.074505}.

\bibitem[Wilson and Zimmermann(1972)]{Wilson_OPE_1972}
K.~G. Wilson and W.~Zimmermann.
\newblock {Operator product expansions and composite field operators in the
  general framework of quantum field theory}.
\newblock \emph{Commun. Math. Phys.}, 24:\penalty0 87--106, 1972.
\newblock \doi{10.1007/BF01878448}.

\end{thebibliography}

\end{document}